\documentclass[twocolumn]{webofc}

\usepackage[varg]{txfonts}   
\usepackage{siunitx}
\usepackage{array}
\usepackage{amsmath}
\usepackage{tikz}
\newcommand{\preliminary}[2][3]{
\begin{tikzpicture}
	\node[anchor=south west,inner sep=0] (murks) at (0,0) {#2};
  \begin{scope}[x={(murks.south east)},y={(murks.north west)}]
    \node [rotate=30,scale=#1,text opacity=0.1]
    at (murks.center) {preliminary};
  \end{scope}
\end{tikzpicture}
}
\usepackage{float}
\usepackage{stfloats}
\begin{document}
\title{$K^+\Lambda$(1405) photoproduction at the BGO-OD experiment}
%
%

\author{\firstname{Georg} \lastname{Scheluchin}\inst{1}\fnsep\thanks{\email{scheluchin@physik.uni-bonn.de}}\and
\firstname{Stefan} \lastname{Alef}\inst{1}\and
\firstname{Patrick} \lastname{Bauer}\inst{1}\and
\firstname{Reinhard} \lastname{Beck}\inst{2}\and
\firstname{Alessandro} \lastname{Braghieri}\inst{13}\and
\firstname{Philip} \lastname{Cole}\inst{3}\and \newline
\firstname{Rachele} \lastname{Di Salvo}\inst{4}\and 
\firstname{Daniel} \lastname{Elsner}\inst{1}\and
\firstname{Alessia} \lastname{Fantini}\inst{4,5}\and
\firstname{Oliver} \lastname{Freyermuth}\inst{1}\and
\firstname{Francesco} \lastname{Ghio}\inst{6,7}\and
\firstname{Anatoly} \lastname{Gridnev}\inst{8}\and \newline
\firstname{Daniel} \lastname{Hammann}\inst{1}\and 
\firstname{J\"urgen} \lastname{Hannappel}\inst{1}\and 
\firstname{Thomas} \lastname{Jude}\inst{1}\and
\firstname{Katrin} \lastname{Kohl}\inst{1}\and
\firstname{Nikolay} \lastname{Kozlenko}\inst{8}\and
\firstname{Alexander} \lastname{Lapik}\inst{9}\and \newline
\firstname{Paolo} \lastname{Levi Sandri}\inst{10}\and 
\firstname{Valery} \lastname{Lisin}\inst{9}\and 
\firstname{Giuseppe} \lastname{Mandaglio}\inst{11,12}\and
\firstname{Roberto} \lastname{Messi}\inst{4,5}\and
\firstname{Dario} \lastname{Moricciani}\inst{4}\and
\firstname{Vladimir} \lastname{Nedorezov}\inst{9}\and \newline
\firstname{Dmitry} \lastname{Novinsky}\inst{8}\and 
\firstname{Paolo} \lastname{Pedroni}\inst{13}\and
\firstname{Andrei} \lastname{Polonski}\inst{9}\and
\firstname{Bj\"orn-Eric} \lastname{Reitz}\inst{1}\and
\firstname{Mariia} \lastname{Romaniuk}\inst{4}\and
\firstname{Hartmut} \lastname{Schmieden}\inst{1}\and \newline
\firstname{Victorin} \lastname{Sumachev}\inst{8}\and
\firstname{Viacheslav} \lastname{Tarakanov}\inst{8}\and
\firstname{Christian} \lastname{Tillmanns}\inst{1}}

\institute{Rheinische Friedrich-Willhelms-Universit\"at Bonn, Physikalisches Institut, Nu\ss allee 12, 53115 Bonn, Germany \and 
Helmholtz-Institut fuer Strahlen- und Kernphysik, Universitaet Bonn, Nussallee 1-16, D-53115 Bonn Germany \and
Lamar University, Department of Physics, Beaumont, Texas, 77710, USA \and
INFN Roma ``Tor Vergata", Rome, Italy \and
Università di Roma ``Tor Vergata", Via della Ricerca Scientifica 1, 00133 Rome, Italy \and
INFN sezione di Roma La Sapienza, P.le Aldo Moro 2, 00185 Rome, Italy \and
Istituto Superiore di Sanita, Viale Regina Elena 299, 00161 Rome, Italy \and
Petersburg Nuclear Physics Institute, Gatchina, Leningrad District, 188300, Russia \and
Russian Academy of Sciences Institute for Nuclear Research, prospekt 60-letiya Oktyabrya 7a, Moscow 117312, Russia \and
INFN - Laboratori Nazionali di Frascati, Via E. Fermi 40, 00044 Frascati, Italy \and
INFN sezione Catania, 95129 Catania, Italy \and
Universita degli Studi di Messina, Via Consolato del Mare 41, 98121 Messina, Italy\and
\,INFN sezione di Pavia, Via Agostino Bassi, 6 - 27100 Pavia, Italy}


\abstract{%
Since the discovery of the $\Lambda(1405)$, it remains poorly described by conventional constituent quark models, and it is a candidate for having an "exotic" meson-baryon or "penta-quark" structure, similar to states recently reported in the hidden charm sector.

The $\Lambda(1405)$ can be produced in the reaction $\gamma p \rightarrow K^+\Lambda(1405)$. The pure I=0 decay mode into $\Sigma^0\pi^0$ is prohibited for the mass-overlapping $\Sigma(1385)$. Combining a large aperture forward magnetic spectrometer and a central BGO crystal calorimeter, the BGO-OD experiment is ideally suited to measure this decay with the $K^+$ in the forward direction. Preliminary results are presented.

*Supported by DFG (PN 388979758, 405882627).
}
\maketitle

\section{Introduction}
\label{s_intro}

For decades hadron spectroscopy has been used to investigate the strong interaction. Experiments on protons showed that baryons consist of three valance quarks and sea-quarks. 
While three quark models show a good agreement between calculated and measured for most of the states, the $\Lambda(1405)$ deviates from the predictions. It is lighter in mass than its non-strange counterpart, $N(1535)$, even thought it has a strange quark in its composition. Furthermore the mass distribution, also called line shape, does not follow a Breit-Wigner distribution. The initial inconsistencies elevated $\Lambda(1405)$ as an unconventional state candidate since the discovery more than 50 years ago.

Nowadays there are more theoretical models for $\Lambda(1405)$ having a molecule like structure of $N\bar{K}$. Lattice QCD calculations give more support for the molecule like structure compared to a genuine three quark state \cite{bib_latticeQCD}\cite{bib_molina}. A study using a chiral unitarity model, where the resonance is generated dynamically from $N\bar{K}$ interactions with other channels constructed from the octets of baryons and mesons, shows that the line shape would depend on the
\begin{figure}[H]
\centering
\includegraphics[width=0.45\textwidth,clip]{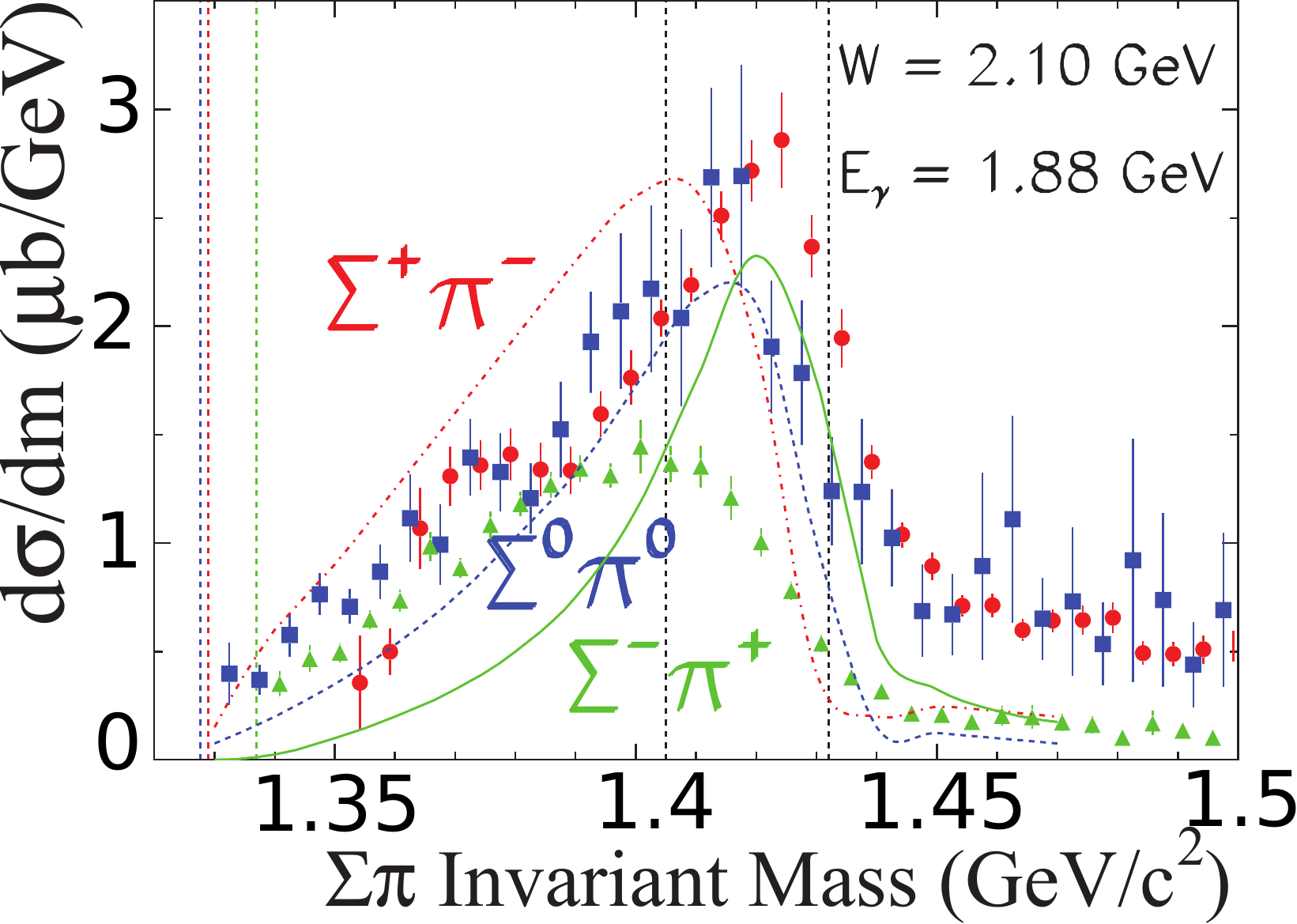}
\caption{Line shape results from the CLAS experiment \cite{bib_CLAS_lineshape}\cite{bib_CLAS_lineshape2} for different decay modes. Colored lines show model predictions of Nacher et al. \cite{bib_Nacher}. Figure taken from \cite{bib_CLAS_lineshape}.}
\label{f_CLAS_lineshape}       
\end{figure}
\begin{figure}[h]
\centering
\includegraphics[width=0.3\textwidth,clip]{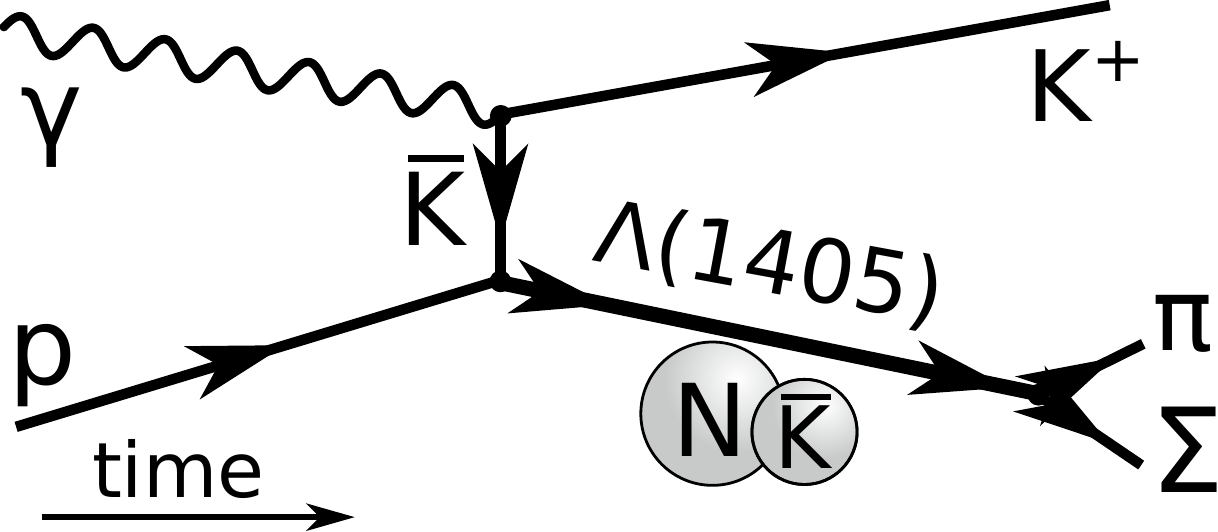}
\caption{Possible photoproduction scheme of the $\Lambda(1405)$.}
\label{f_tproccess}       
\end{figure}
\noindent
decay mode \cite{bib_Nacher}. The results from the CLAS experiment are seen in figure~\ref{f_CLAS_lineshape} \cite{bib_CLAS_lineshape}. The measurements show that the line shape does differ for the decay modes. On close inspection it is visible that the measurement and predictions for the charged decays differ from the predictions and a second experiment could help to resolve this discrepancy.

\begin{figure*}[b]
\centering
\includegraphics[width=0.8\textwidth,clip]{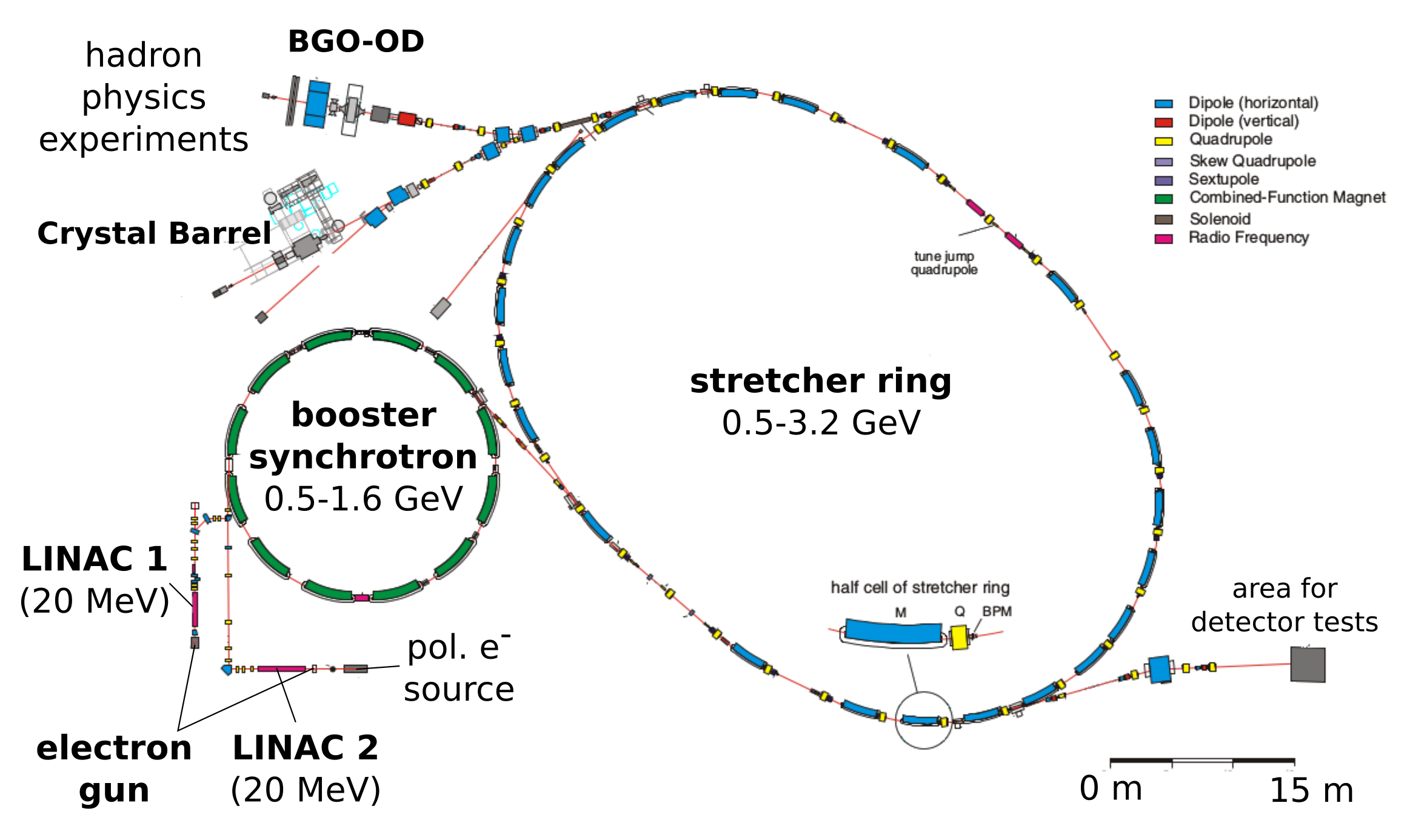}
\caption{Overview of the Electron Stretcher Accelerator \cite{bib_elsa} in Bonn, showing the main components.}
\label{f_elsa}       
\end{figure*}

$\Lambda(1405)$ can be produced via $\gamma p \rightarrow K^+\Lambda(1405)$ as seen in figure \ref{f_tproccess}. With the assumption of a molecular like state, one can assume that the cross section for $\Lambda(1405)$ is increased if the transferred momentum to the baryon is low. This means that the $K^+$ needs to take most of the photon momentum, which corresponds to extreme forward angles in a fixed target experiment, while the $\Lambda(1405)$ decays almost at rest isotropically. The BGO-OD\footnote{\textbf{B}ismuth \textbf{G}ermanum \textbf{O}xide calorimeter with a  \textbf{O}pen \textbf{D}ipole magnet spectrometer} experiment with its central calorimeter and forward spectrometer is ideally suited to measure such kinematics, which were not yet explored by other experiments for the $\Sigma^0\pi^0$ decay. In the following chapters the experimental setup and particle identification are described in more detail. 
The experiment is explained in section~\ref{s_bgood} and the reaction identification in section~\ref{s_reconst}. Preliminary results are shown in section~\ref{s_results}.

\begin{figure*}[h]
\centering
\includegraphics[width=0.8\textwidth,clip]{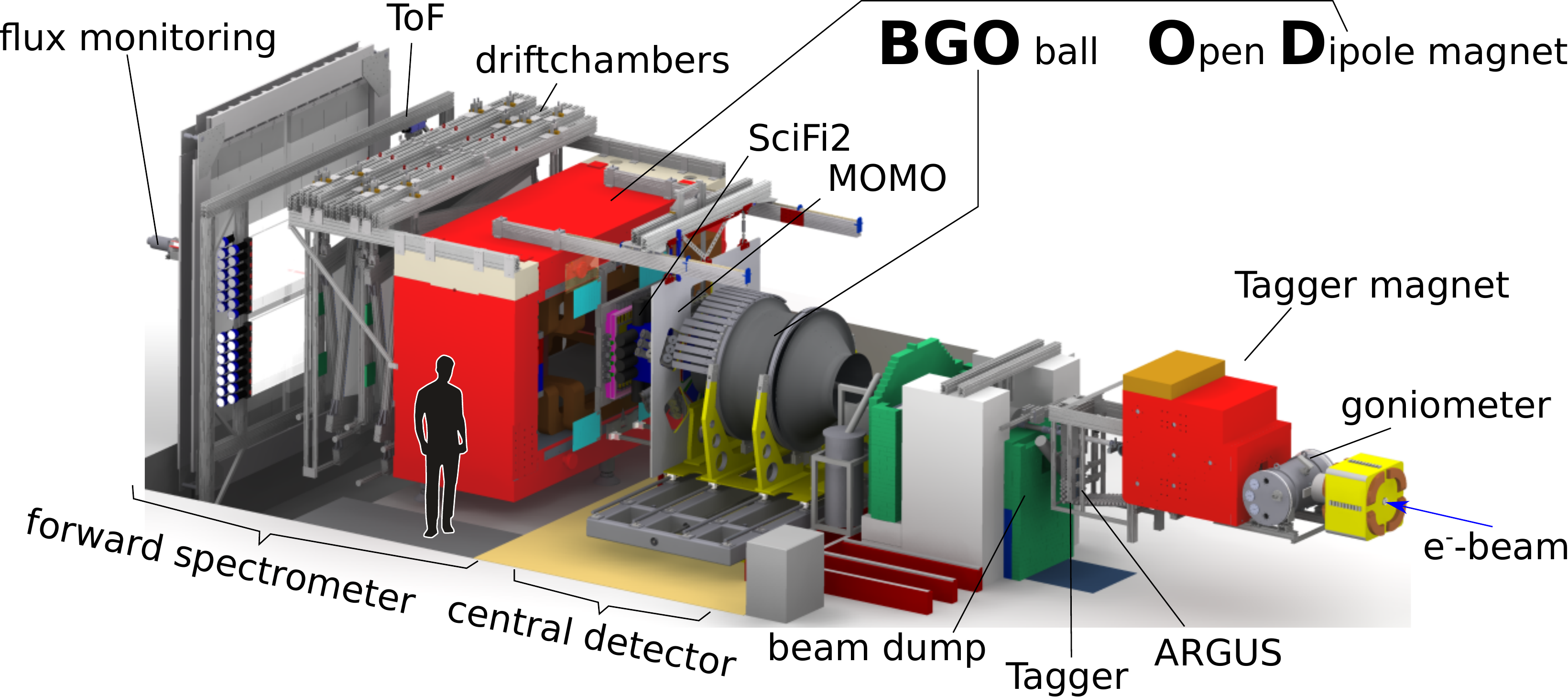}
\caption{Overview of the BGO-OD experiment.}
\label{f_bgood}       
\end{figure*}

\section{The BGO-OD experiment}
\label{s_bgood}
The BGO-OD experiment is located at the ELSA\footnote{The \textbf{E}lectron \textbf{S}tretcher \textbf{A}ccelerator in Bonn(Germany)} facility in Bonn. ELSA is an electron accelerator with a quasi continuous beam up to \SI{3.2}{GeV} \cite{bib_hillert}. In figure \ref{f_elsa} the general overview of the accelerator facility is shown. It is a three stage electron accelerator. Using a linear accelerator LINAC the electrons are accelerated to relativistic velocities before the injection into the booster synchrotron. Here the electrons are accelerated to an energy of \SIrange{0.5}{1.6}{GeV} before they are guided to the stretcher ring. The stretcher ring needs to be filled multiple times by the synchrotron to be completely full. After this the electrons are accelerated up to \SI{3.2}{GeV}. The electrons are only extracted partially to either the CBELSA/TAPS\cite{bib_cb} or the BGO-OD experiment each revolution. This results in a quasi continuous electron beam for the duration of \SIrange{5}{12}{s} until the ring is empty and the process starts anew.

\begin{figure}[h]
\centering
\includegraphics[width=0.45\textwidth,clip]{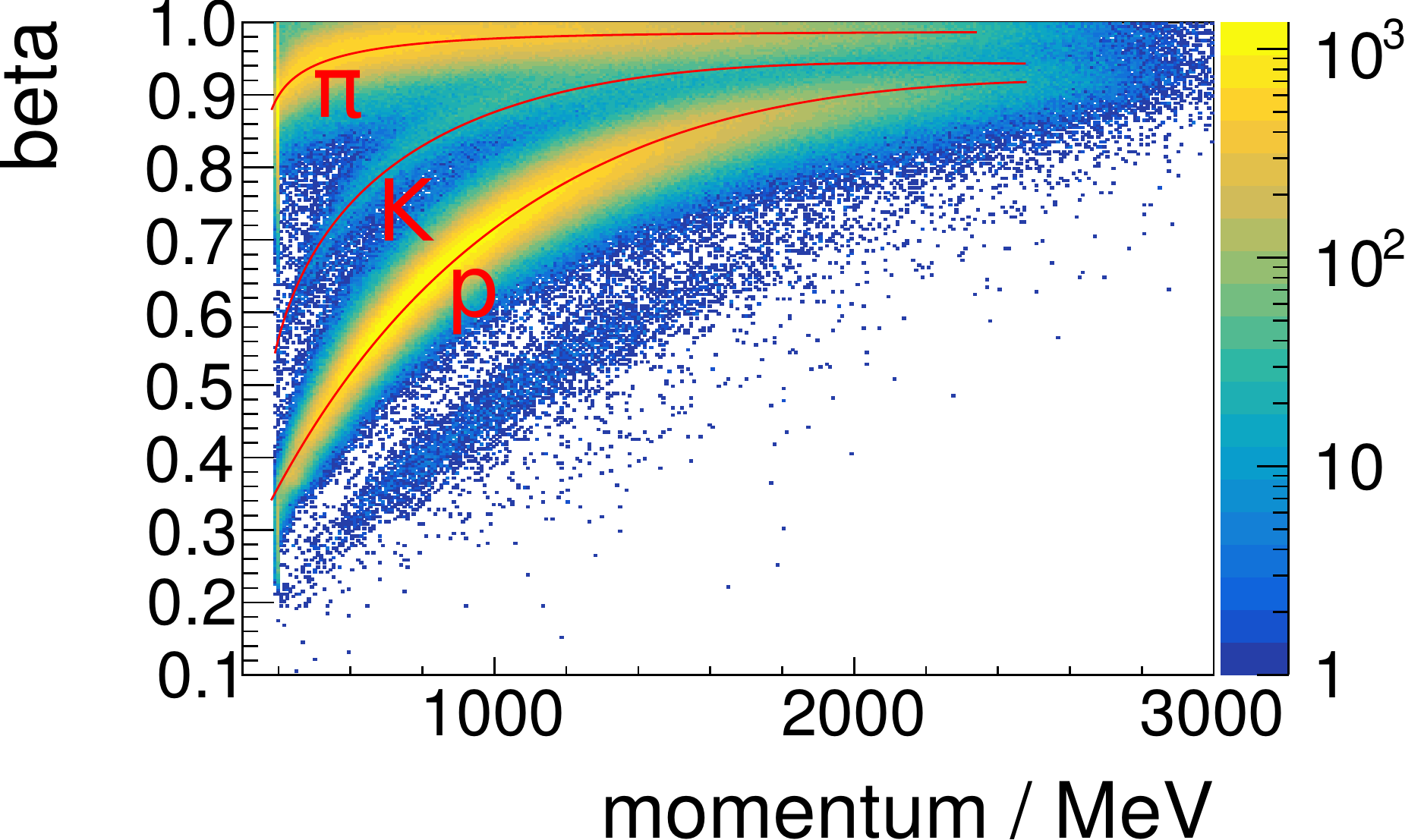}
\caption{Particle velocity $\beta$ against momentum $p$ detected in the forward spectrometer. Due to the mass difference the charged particles $\pi$, $K$ and $p$ can be distinguished indicated by lines. }
\label{f_betavsmomentum}       
\end{figure}

In figure \ref{f_bgood} the BGO-OD experiment is shown in more detail. The electron beam hits a radiator inside the goniometer tank, which creates a real photon beam via bremsstrahlung. The energy of the photons is determined by measurement of the bremsstrahlung electron with the Tagger detector. The photon beam interacts with the target cell inside the BGO ball, which is filled with liquid hydrogen or deuterium. The final state particles of the reaction are detected with the BGO calorimeter using bismuth germinate oxide crystals between polar angles \SIrange{25}{155}{\degree}. Charged particles traveling in $\theta<$\SI{12}{\degree} are detected in the forward detector. The track trajectory before the open-dipole magnet is measured with the MOMO and SciFi2 scintillating fibre detectors, and the drift chambers measure the trajectory after the magnet. The measured track curvature is used to determine the momentum, while the ToF walls at the end of the experiment measure the velocity via time of flight. Both measured quantities can be used for particle identification as seen in figure \ref{f_betavsmomentum}.

\section{$\Lambda(1405)$ identification}
\label{s_reconst}
$\Lambda(1405)$ was identified via the $\Sigma^0\pi^0$ decay, which is prohibited for $\Sigma(1385)$. The complete reactions is $\gamma p \rightarrow K^+ \Lambda(1405) \rightarrow K^+ \Sigma^0 \pi^0$. The reaction can be identified by detecting $K^+$ and $\pi^0$, while the $\Sigma^0$ is identified via missing mass techniques. This can be achieved with the $K^+$ detected in the forward spectrometer and is described in section \ref{s_kforward}. An additional technique is used by detecting the full final state increasing the $K^+$ polar angle acceptance in exchange for lower statistics, which is described in section \ref{s_fulltop}.

\subsection{$K^+$ at extreme forward angles}
\label{s_kforward}
\begin{figure}[h]
\centering
\includegraphics[width=0.45\textwidth,clip]{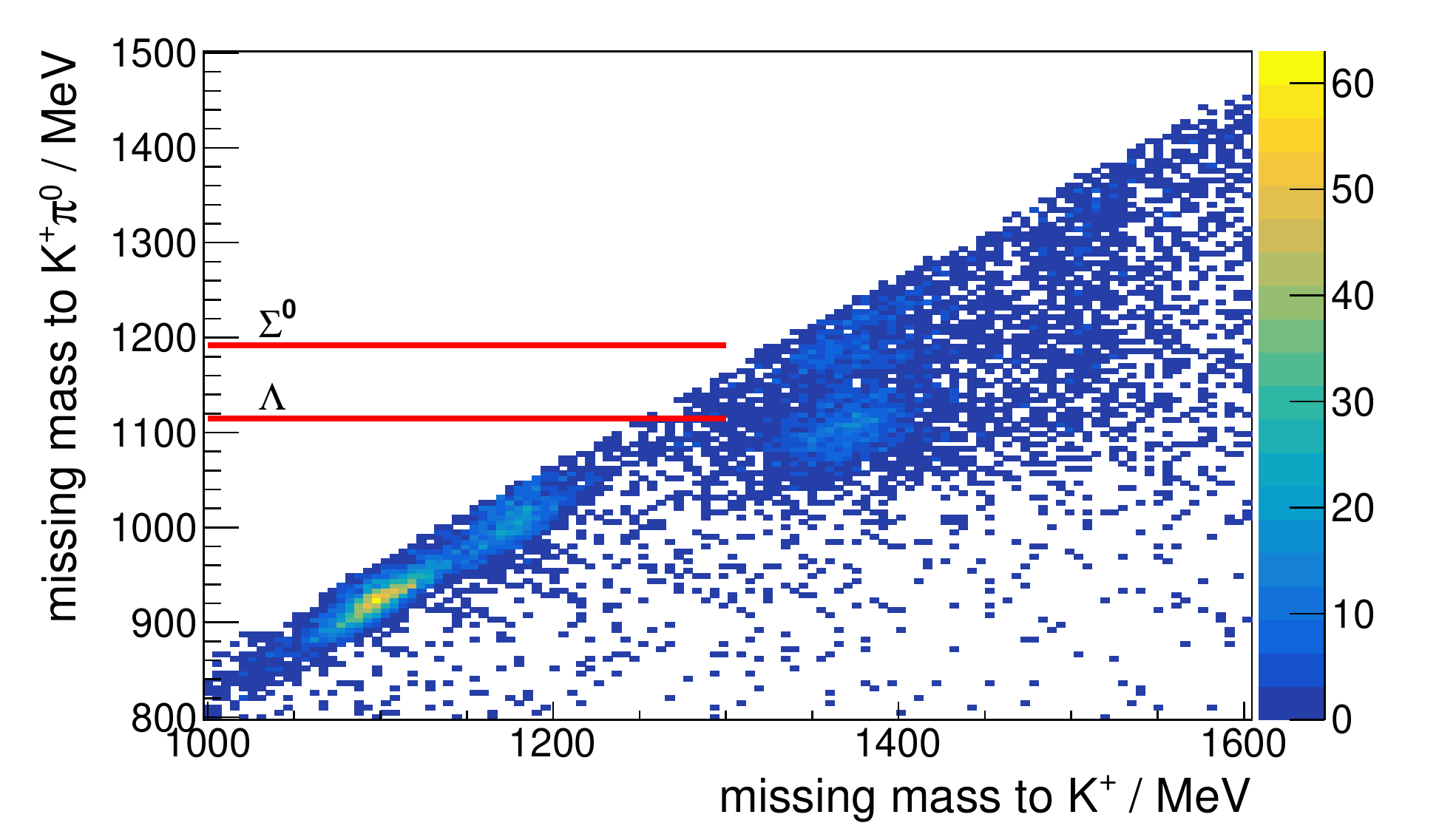}
\caption{Missing mass to $K^+\pi^0$ against missing mass to $K^+$, while the $K^+$ was detected in the forward spectrometer.}
\label{f_forwardmass}       
\end{figure}

\begin{figure}[h]
\centering
\includegraphics[width=0.45\textwidth,clip]{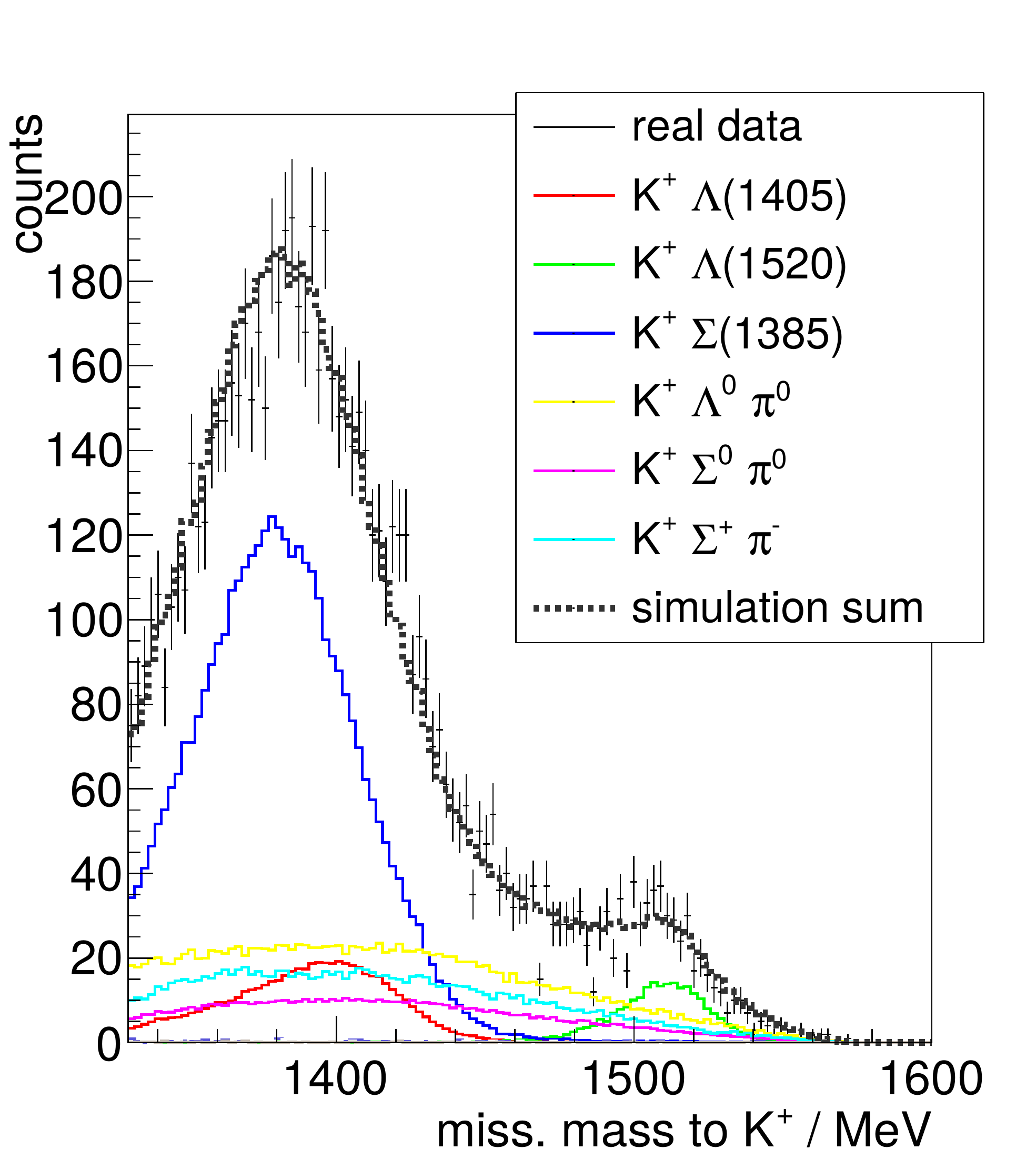}
\caption{Missing mass to $K^+$. This is a projection of the two dimensional histogram in figure \ref{f_forwardmass}. The colored lines show the contributions of different simulated channels fitted to the data.}
\label{f_forwardmass_proj}       
\end{figure}

The $K^+$ can be identified in the forward spectrometer using the measured velocity and momentum of the charged particles, as seen in figure \ref{f_betavsmomentum}. A $\pi^0$ can be identified by combining two measured photons in the central calorimeter and select events with invariant mass close to the $\pi^0$ mass. This leaves the $\Sigma^0$ particle to be identified through missing mass techniques. Figure \ref{f_forwardmass} shows the missing mass to $K^+$ and $K^+\pi^0$ systems. The red lines indicate the possible missing hyperons in the reaction. The events with a $\Lambda$ mass originate from the prominent $\Sigma(1385)^0\rightarrow\Lambda\pi^0$ decay. While it looks like there are some events with missing $\Sigma^0$, studies showed that most of these events are combinatorial background from $\Sigma(1385)$ as seen in figure \ref{f_forwardmass_proj}. 
First preliminary results for the differential cross section can be seen in section \ref{s_results_crosssection}. Further analysis is needed to improve the signal to background ratio.

\subsection{Full topology measurement}
\label{s_fulltop}

\begin{figure}[h]
\centering
\includegraphics[width=0.45\textwidth,clip]{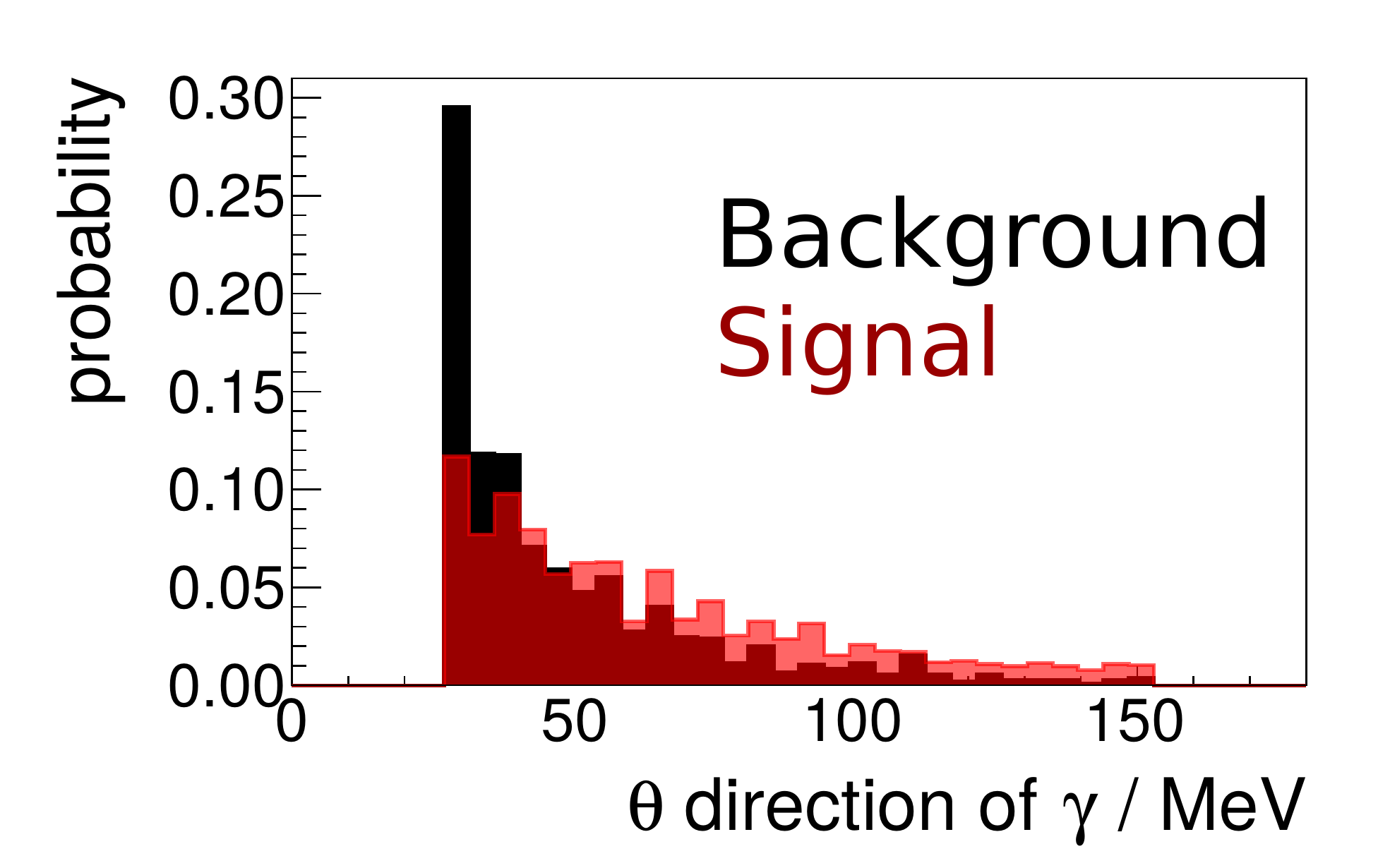}
\caption{Angular distribution of photon compared between signal in red and background in black. }
\label{f_background}       
\end{figure}

\begin{figure}[h]
\centering
\includegraphics[width=0.45\textwidth,clip]{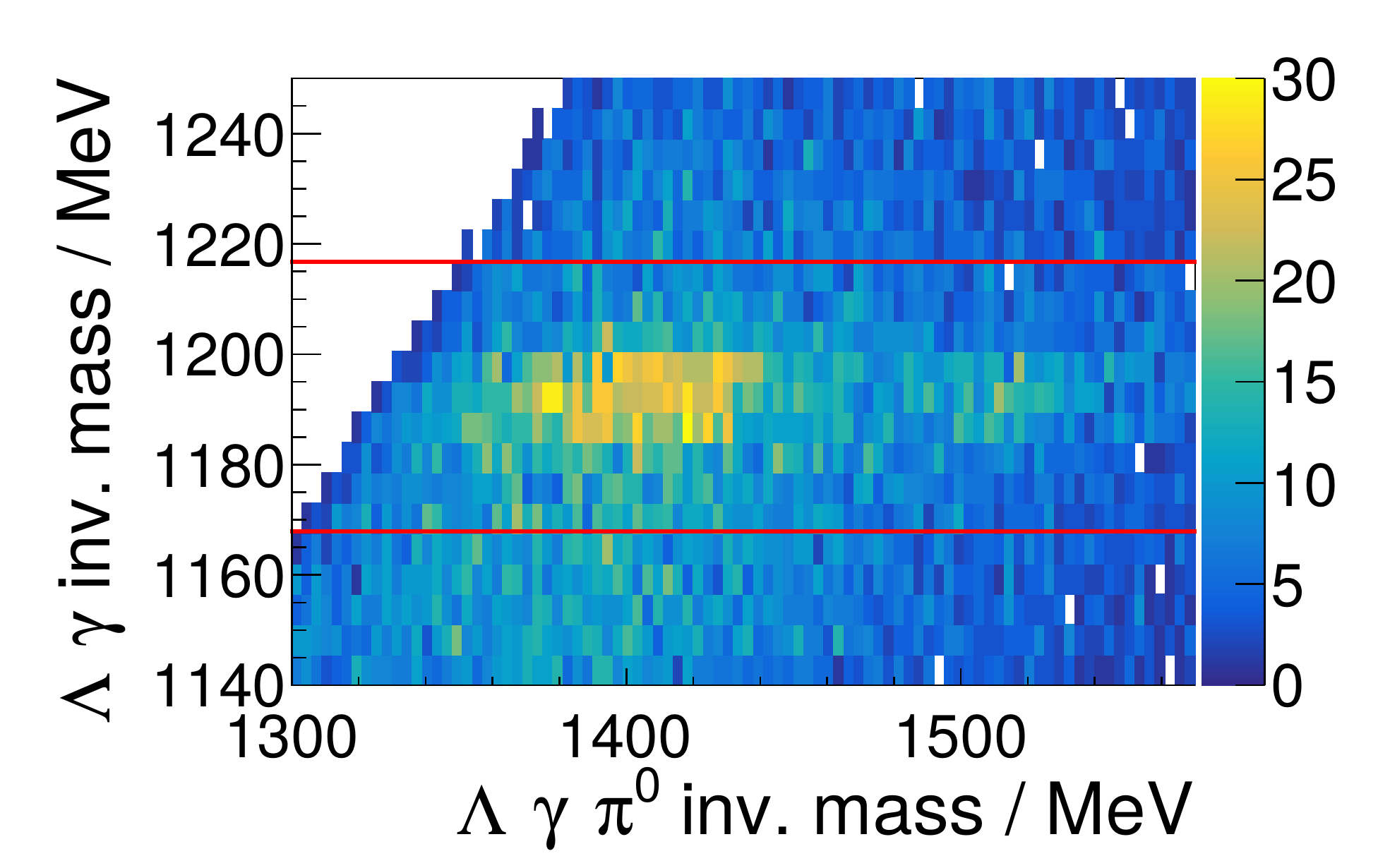}
\caption{Invariant mass of $\gamma\Lambda$ against $\gamma\Lambda\pi^0$ in the reaction $\gamma p \rightarrow \Sigma^0\pi^0K^+$. }
\label{f_2d_fulltop}       
\end{figure}

\begin{figure}[h]
\centering
\includegraphics[width=0.45\textwidth,clip]{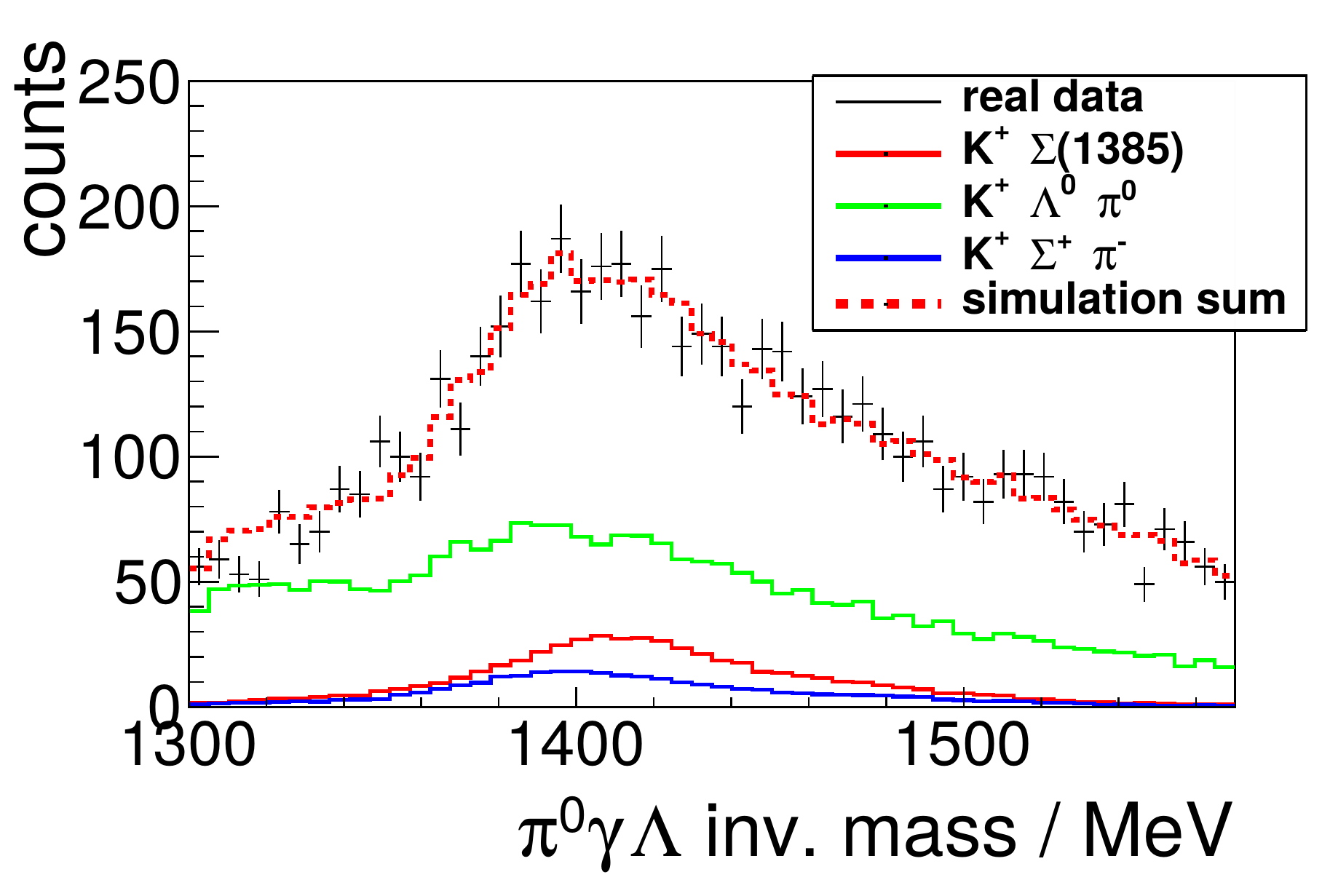}
\caption{$\Lambda\gamma\pi^0$ mass projection of figure \ref{f_2d_fulltop}, without events where $\Lambda\gamma$ mass is close to $\Sigma^0$ mass. The colored lines show example reactions contributing to the background shape, while the red dotted line shows the total fit including also reactions not shown in figure. }
\label{f_proj_sigmapi}       
\end{figure}

\begin{figure}[h]
\centering
\includegraphics[width=0.45\textwidth,clip]{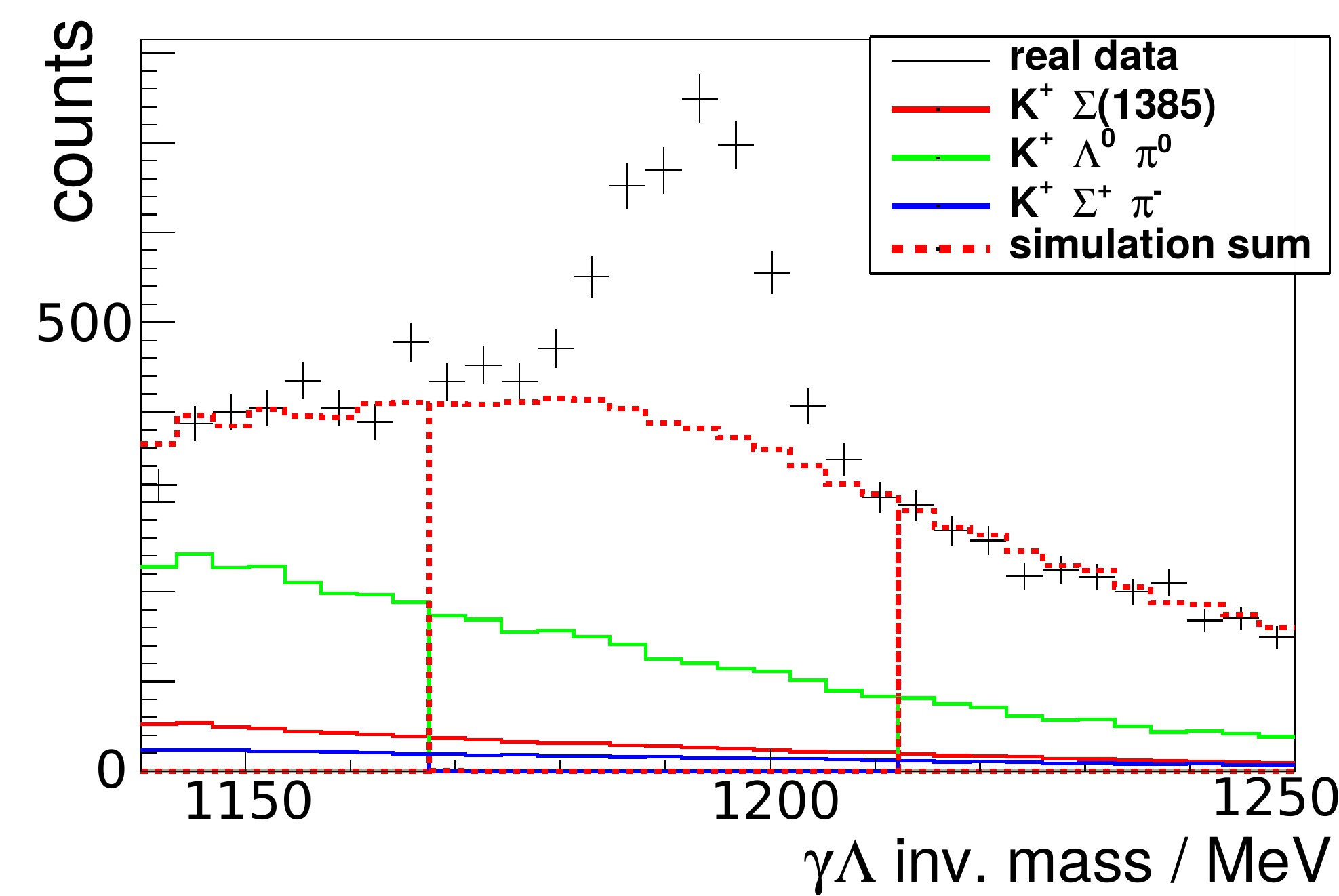}
\caption{$\Lambda\gamma$ mass projection of figure \ref{f_2d_fulltop}. The colored lines show example reactions contributing to the background shape, while the red dotted line shows the total fit including also reactions not shown in figure. }
\label{f_poj_sigma}       
\end{figure}

\begin{figure}[h]
\centering
\includegraphics[width=0.45\textwidth,clip]{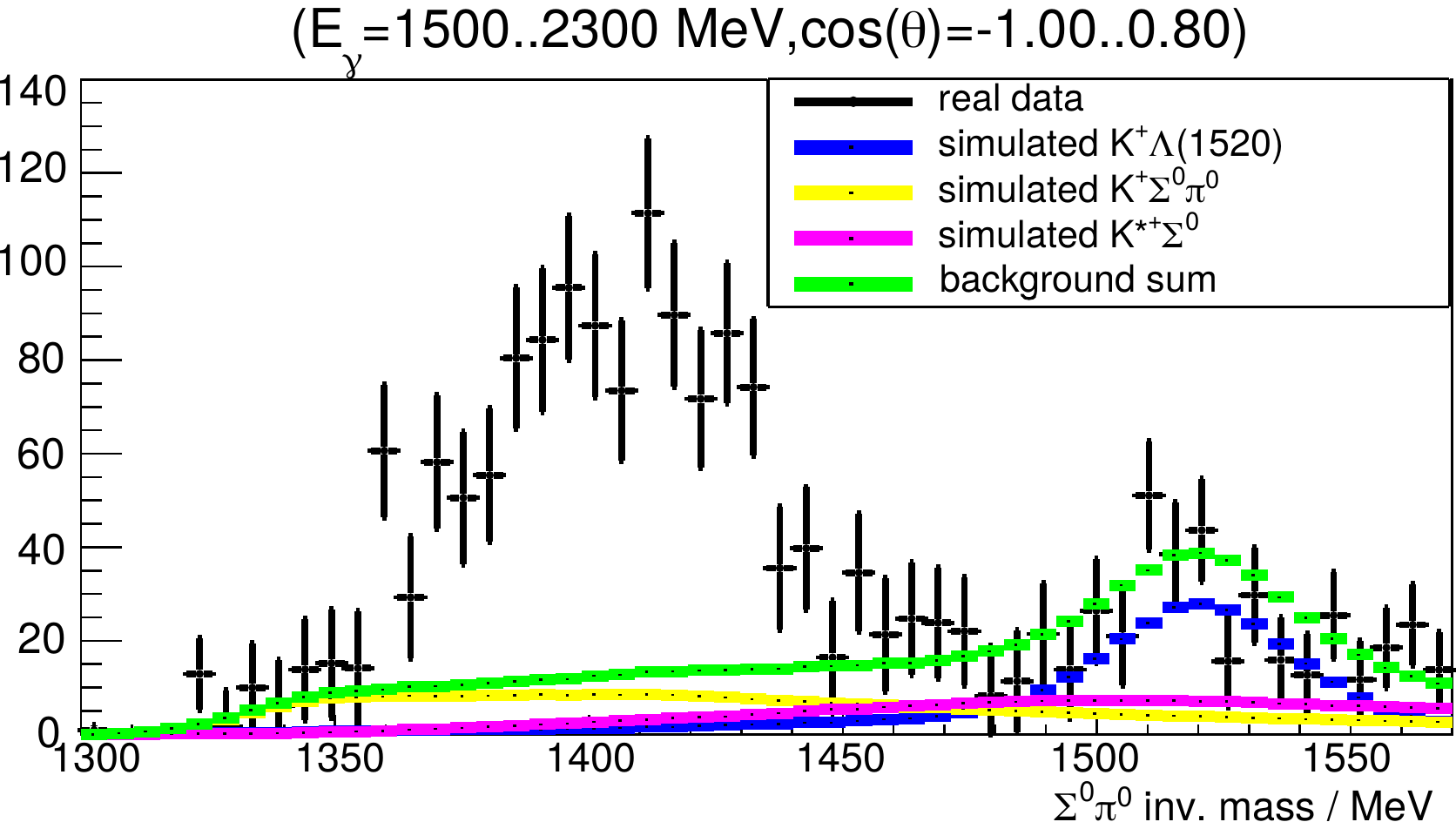}
\caption{Invariant mass of $\Sigma^0\pi^0$ after all analysis steps. The colored lines show the estimation for not $ \Lambda(1405)$ reactions.}
\label{f_backgourndfree}       
\end{figure}

Measuring all final state particles in the reaction:
\begin{eqnarray}
 \gamma p  & \rightarrow &  \Lambda(1405) K^+  \\
             & \rightarrow & \Sigma^0 \pi^0 K^+ \rightarrow \Lambda \gamma \pi^0 K^+ \\
             & \rightarrow & p \pi^- \gamma \gamma \gamma K^+ ,
\end{eqnarray}
allows identification with $K^+$ polar angles outside the forward spectrometer acceptance. 
However the lack of particle identification increases the number of combinatorial background. Therefore after creating all possible combinatorics, a kinematics fit is used to improve the mass resolution and reduce background via confidence level selection cuts. In figure \ref{f_background} the angular distribution of the photons is shown for signal and background simulation after the kinematic fit. The combinatorial background passes the selection cuts predominantly with low photon angles. Events with low photon angles are removed from the \mbox{analysis} to improve the signal to noise ratio at the cost of statistics. The $\gamma\Lambda$ against $\gamma\Lambda\pi^0$ invariant mass distribution is plotted in figure \ref{f_2d_fulltop}. The signal and background can be distinguished, as only the signal shows a correlation to the $\Sigma^0$ mass in the $\Lambda\gamma$ invariant mass. A peak for $\Lambda(1405)$ and $\Lambda(1520)$ is visible as both decay to $\Sigma^0\pi^0$. To subtract the background, the mass outside the signal region, marked by the red lines is used to modulate the amplitude of different simulated reactions, for example $K^+\Sigma(1385)$. In figure \ref{f_proj_sigmapi} the projection of the $\Lambda\gamma\pi^0$ mass is shown, where the signal with $\Lambda\gamma$ mass close to $\Sigma^0$ is subtracted. With this fit the background distribution is modeled and can be extrapolated to the region of the signal. In figure \ref{f_poj_sigma}, the $\Lambda\gamma$ projection of the two dimensional fit is shown. The signal region was not included in the fit. Subtracting the fitted background distribution from the data yields the invariant mass distribution for pure $\Sigma^0\pi^0$ reactions, which is plotted in figure \ref{f_backgourndfree}. The contribution of non $\Lambda(1405)$ events is estimated by the green line. At this point the line shape and differential cross section can be extracted by subtraction of the simulated contributions. The preliminary results are shown in section \ref{s_results}.

%

\section{Preliminary results for $\Lambda(1405)\rightarrow\Sigma^0\pi^0$ line shape and differential cross section}
\label{s_results}
Using the analysis steps from the previous section the line shape and differential cross section was determined. For now only the $\Sigma^0\pi^0$ decay was analyzed, as the BGO-OD setup is well suited to measure the decay particles. In section \ref{s_results_lineshape} the line shape results are shown, while section \ref{s_results_crosssection} shows the differential cross section.  

\subsection{Line shape}
\label{s_results_lineshape}

\begin{figure}[h]
\centering
\preliminary{
\includegraphics[width=0.45\textwidth,clip]{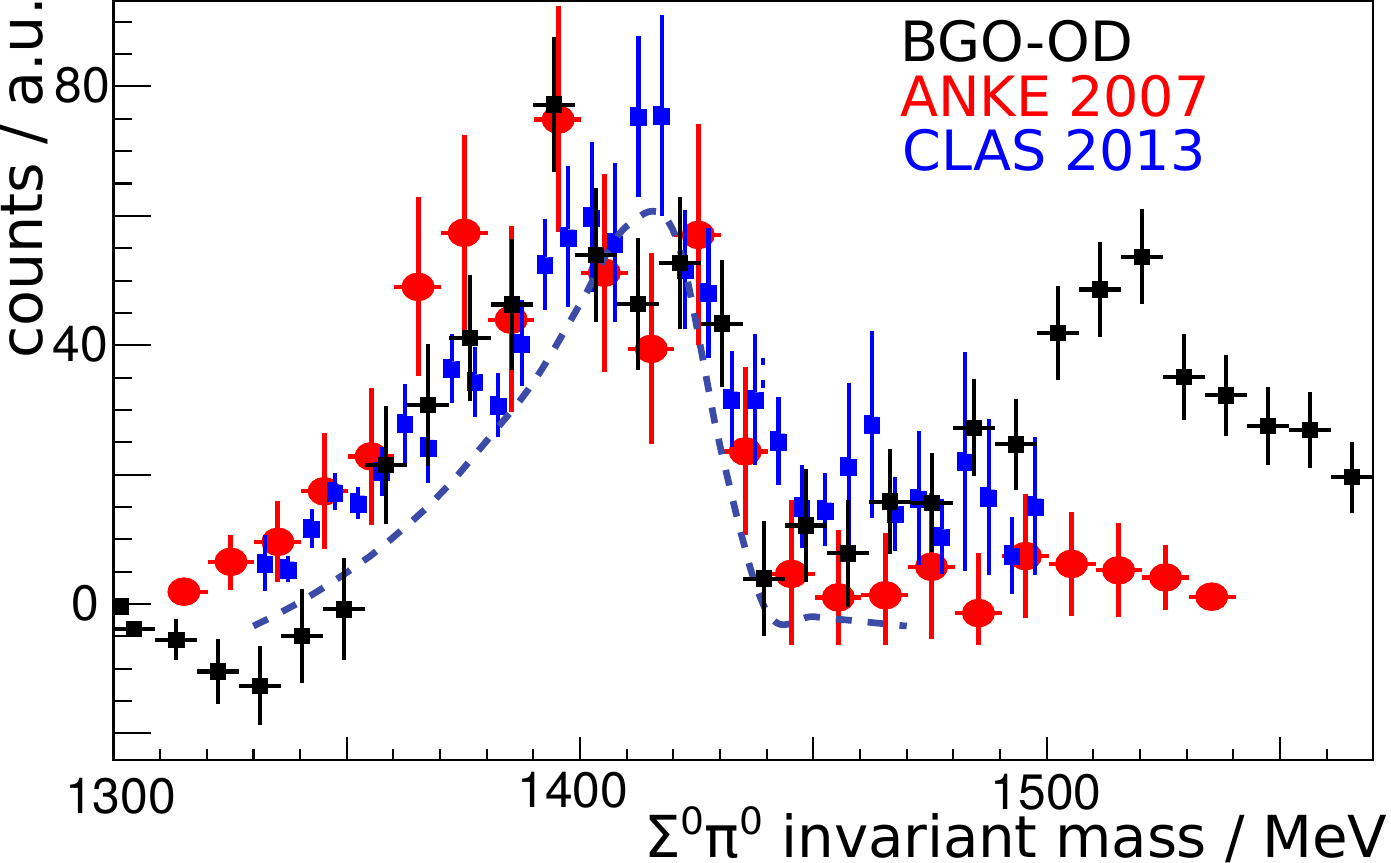}}
\caption{Preliminary line shape of $\Lambda(1405)$ for $E_\gamma=1500$~to~$2300$~MeV compared to the CLAS experiments in blue\cite{bib_CLAS_lineshape2} and ANKE in red\cite{bib_anke}. The dashed blue line marks the predictions by J.C.Nacher\cite{bib_Nacher}. The BGO-OD and ANKE results are in yield with no normalization. To compare the line shape, the results of the experiments were scaled such that the maximum values agree.  \newline  }
\label{f_lineshapes}       
\end{figure}

The preliminary results on the line shape are compared to other experimental results in figure \ref{f_lineshapes}. The $\Lambda(1520)$ signal was not subtracted for consistency checks. The results are statistically comparable to the CLAS experiment, which derives from the better acceptance for this decay mode compared to the CLAS setup. Comparing the ANKE and BGO-OD results indicates a double peak structure in the line shape, which agree to the two pole structure of $\Lambda(1405)$\cite{bib_doublepeak}. This seems not to be present in the CLAS data and  the statistical error is too big to make a definite statement. All experimental results agree statistically to the predicted line shape in the figure. As mentioned in the introduction the deviations to the predictions show up in the charged decay modes, which will be analyzed in the future.

\subsection{Differential cross section}
\label{s_results_crosssection}

\begin{figure*}[h]
\centering

\preliminary{
\includegraphics[width=0.70\textwidth,clip]{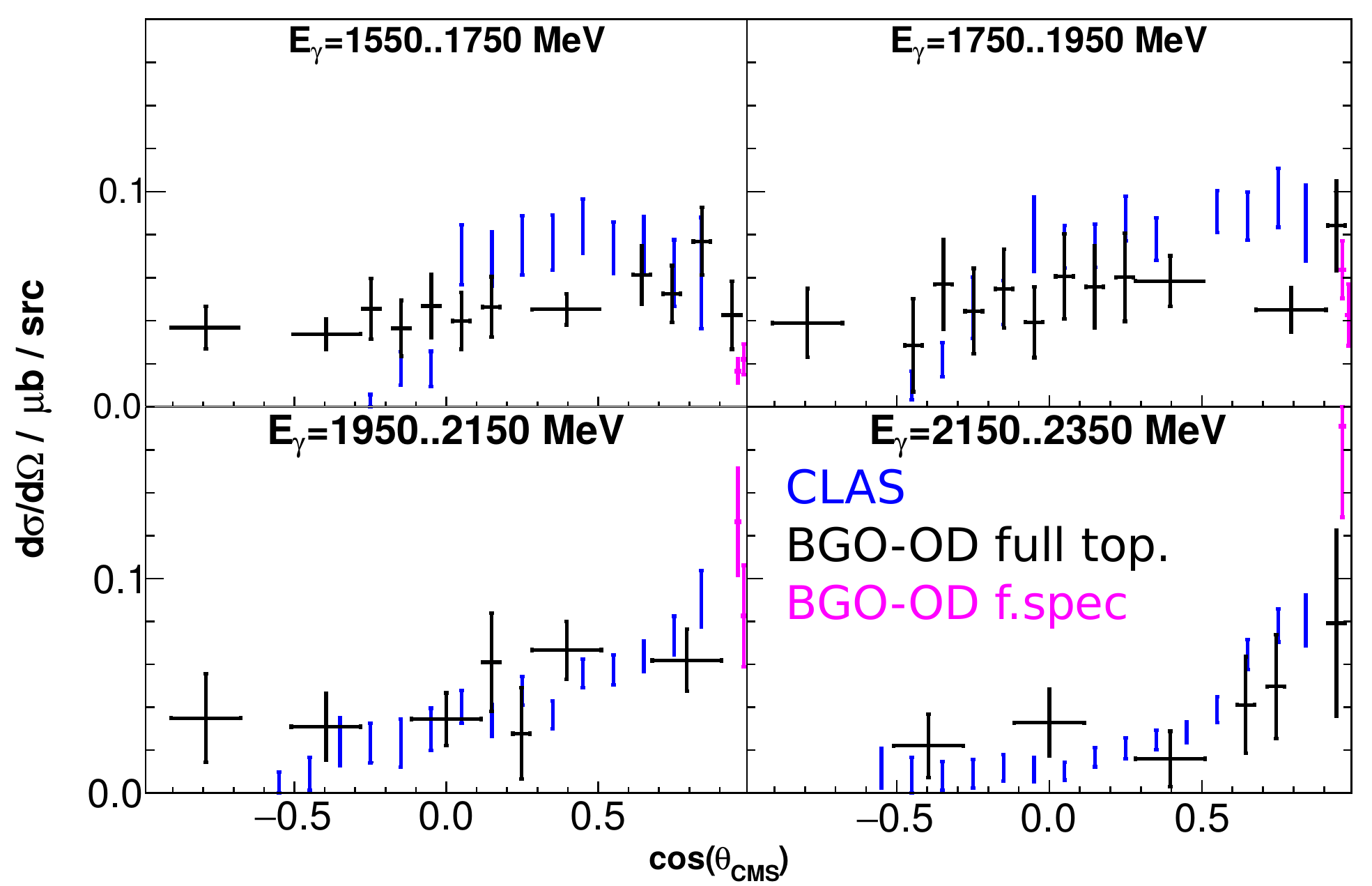}}
\caption{Preliminary differential cross section against angle compared to the CLAS experiment\cite{bib_CLAS_lineshape2} in blue. The black points mark the full topology reconstruction while magenta shows the $K^+$ in the forward spectrometer analysis. }
\label{f_crosssection}       
\end{figure*}

The results on the cross section are plotted in figure \ref{f_crosssection}. In general the cross section agrees to the CLAS results. The forward spectrometer analysis extends the CLAS results to extreme forward angles, which correlates to minimum transfer momentum. While the results agree for higher energies, it suggest the photon energy bin $E_\gamma~=~1550..1750$~MeV shows a more flat distribution compared to the CLAS results. The charged decay modes results of CLAS also show this behavior compared to the their measured neutral decay mode. Further investigation is needed to clarify this.

\section{Summary and Outlook}
\label{s_summary}

The BGO-OD experiment is ideally suited to investigate the formation of unconventional states. Since more than 50 years $\Lambda(1405)$ is a potential candidate for such a state. Models describe $\Lambda(1405)$ as a $N\bar{K}$ molecule-like structure, which is formed below the free $N\bar{K}$ production threshold. This is also used as an explanation of the sudden cut-off in the invariant mass distribution (line shape) above \SI{1426}{MeV} as this marks the free production threshold. Thus the line shape is of great interest for the study of $\Lambda(1405)$. Preliminary results for the decay mode $\Lambda(1405)\rightarrow\Sigma^0\pi^0$ line shape and differential cross section agree within statistics to the CLAS experiment, while the differential cross section could be extended to extreme forward angles thanks to the forwards spectrometer of the BGO-OD experiment. The charged decay modes of $\Lambda(1405)$ will be investigated in the future.

\subsection*{Acknowledgements}
I thank the ELSA group for operating and maintaining of the electron accelerator. For the strong support on the maintenance and improvement of our experimental setup I thank the technical staff of the contributing institutions. 

This work was supported by the Deutsche Forschungsgemeinschaft project numbers 388979758 and 405882627.  Our Russian collaborators thank the Russian Scientific Foundation (grant RSF number 19-42-04132) for financial support.

%

\begin{thebibliography}{}
 %
%
\bibitem{bib_latticeQCD}
J.M.M. Hall et al., Physical Review Letters \textbf{114}, doi:10.1103/PhysRevLett.114.132002 (2015)
\bibitem{bib_molina}
R. Molina and M. Döring, Physical Review Letters D \textbf{94},056010 (2016)
\bibitem{bib_Nacher}
J.C.Nacher et al., Physical Review Letters B \textbf{455},55 (1999)
\bibitem{bib_CLAS_lineshape}
K.Moriya et al.(CLAS Collaboration), Physical Review Letters C \textbf{87},035206 (2013)
\bibitem{bib_CLAS_lineshape2}
K.Moriya et al.(CLAS Collaboration), Physical Review Letters C \textbf{88},045201 (2013)

\bibitem{bib_hillert}
W. Hillert, Europ. Phys. Jour. A \textbf{28}, 139 (2006)
\bibitem{bib_elsa}
On basis of website. url: http://www-elsa.physik.uni-bonn.de/index\_en.html 
\bibitem{bib_cb}
M.Nanova et al. (CB-ELSA Collaboration), Phys. Rev. C, \textbf{94} (2016)
\bibitem{bib_anke}
I.Zychor et al., Physical Review Letters B \textbf{660}, 167-171 (2008)
\bibitem{bib_doublepeak}
J.A. Oller and U.-G. Meißner, Phys. Lett. B \textbf{500}, 263-272 (2001)  



\end{thebibliography}
%
%

\end{document}